\documentstyle[12pt,epsf]{article}

\renewcommand{\a}{\alpha}
\renewcommand{\b}{\beta}

\renewcommand{\d}{\delta}
\newcommand{\D}{\Delta} 

\newcommand{\ep}{\epsilon}
\newcommand{\G}{\Gamma}

\newcommand{\la}{\lambda}

\newcommand{\si}{\sigma}
\newcommand{\Si}{\Sigma}

\newcommand{\zb}{\bar z}

\newcommand{\p}{\partial}
\renewcommand{\pb}{\bar \p}

\newcommand{\non}{\nonumber\\}

\newcommand{\lb}{\left(}
\newcommand{\rb}{\right)}
\renewcommand{\le}{\left\lbrack}
\newcommand{\re}{\right\rbrack}

\newcommand{\beq}{\begin{equation}}
\newcommand{\eeq}{\end{equation}}
\newcommand{\beqa}{\begin{eqnarray}}
\newcommand{\eeqa}{\end{eqnarray}}

\newcommand{\refeq}[1]{(\ref{#1})}
\newcommand{\postscript}[2]
 {\setlength{\epsfxsize}{#2\hsize}
  \centerline{\epsfbox{#1}}}
\def\half{{\mbox{\small  $\frac{1}{2}$}}}

\newcommand{\prl}[1]{{  Phys.~Rev.~Lett.}~{\bf {#1}}}
\newcommand{\plb}[1]{{  Phys.~Lett.}~{\bf {#1B}}}
\newcommand{\pla}[1]{{  Phys.~Lett.}~{\bf {#1A}}}
\newcommand{\npb}[1]{{  Nucl.~Phys.}~{\bf B{#1}}}

\newcommand{\ann}[1]{{  Ann.~Phys.~(N.Y.)}~{\bf {#1}}}

\def\vN{{\vec N}}

\def\vn{{\vec n}}
\def\vnp{{\vec n}_+}
\def\vnm{{\vec n}_-}
\newcommand{\vX}[1]{\p_{#1}{\vec X}}
\def\ab{{\bar a}}
\def\bb{{\bar b}}

\begin{document}
\baselineskip 20pt plus 2pt
\begin{titlepage}
\renewcommand{\thefootnote}{\fnsymbol{footnote}}
\begin{flushright}
\parbox{1.5in}
{
 UW-IFT-8/96\\
 April, 1996\\}
\end{flushright}
\vspace*{.5in}
\begin{centering}
{\Large New rigid string instantons in $R^4$\footnote{ 
Work supported, in part, 
by Polish State Committee for Scientific Research (KBN).
}}\\
\vspace{2cm}
{\large        Jacek Pawe\l czyk}\\
\vspace{.5cm}
        {\sl Institute of Theoretical Physics, Warsaw University,\\
        Ho\.{z}a 69, PL-00-681 Warsaw, Poland.}\\
\vspace{.5in}
\end{centering}
\begin{abstract}
New rigid string instanton equations are derived. 
Contrary to standard case, the equations split into three families.
Their solutions in $R^4$ are discussed and explicitly presented in some cases.
\end{abstract}

\end{titlepage}

\setcounter{section}{-1}

\section{Introduction}

It is known that perturbatively rigid string
\cite{polrig,rig} 
is trivial in the sense that at low energies it is equivalent to
the Nambu-Goto string. There is a hope that non-perturbative effects may change
this behaviour. In fact, lattice simulations indicate appearance of a
non-perturbative IR 
fixed point for the 3d rigid string \cite{amb}. For higher dimensional target
spaces the situation is unclear.  
One of the non-perturbative signature of field theory models are instantons. 
Certain instanton equations of rigid string appeared for 
the first time in \cite{polrig} and then their solutions and 
properties were discussed in \cite{wheater,rob}. These instantons appeared to
be non-compact surfaces in $R^4$ and as we shall show they are somehow
exceptional examples of more general instantons.  
One can also find some
remarks about rigid string instantons in \cite{other}. Despite these works not
much have been established toward classifications of instantons and their 
relevance for string dynamics. 

In this note we are going to investigate rigid string instantons in $R^4$ more
throughly. In particular we show that 
equations of  \cite{polrig} give instantons of very limited
type - our construction yield much bigger family. The instantons are 
classified 
by two topological invariants: the Euler characteristic $\chi$ of the immersed
(closed) Riemann surface $\Si$ and the self-intersection number $I$ of the 
immersion.
The constructed set of instantons is rich enough to cover all
possible values of $\chi,I$. It is interesting to note that, contrary to
ordinary instantons, the rigid string
instantons split into three families. The intersection of these families is 
non-trivial and, except one case, is equivalent to the instantons of
\cite{polrig}.  

Let us recall some basic facts about the rigidity (sometimes called  
extrinsic curvature).
The action is given by \refeq{extc1} and it is known to be 
plagued with plethora of identities 
which allow to exhibit its different aspects.
\beq
\int_\Si\sqrt{g}g^{ab}\p_a t^{\mu\nu}\p_b
t^{\mu\nu}=2\int_\Si\sqrt{g}K_a^{ib}K_b^{ia}=
2\int_\Si\sqrt{g}(\D{\vec X})^2-8\pi \chi.   
\label{extc1}
\eeq
In the above $t^{\mu\nu}\equiv \ep^{ab}\p_aX^\mu \p_b X^\nu/\sqrt{g}$ is the
element 
of the Grassman manifold $G_{4,2}$. Throughout the paper we shall exclusively use 
the induced metric 
$g_{ab}\equiv  \vX{a}\vX{b}$. The tension  tensor $K_{ab}^{i}$ is defined by
the relation:
$\p_a\vX{b}=\G_{ab}^{c}\vX{c}+K_{ab}^{i}\vN_i$, where 
$\vN_i$ ($i$=1,2) are two vectors normal to
the immersed surface. The Euler
characteristic of the Riemann surface $\Si$ is given by Gauss-Bonnet formula
$\chi=\frac{1}{4\pi}\int_\Si\sqrt{g}R$.  In the course of the paper we shall
heavily use identities
expressing  $\chi$ and the self-intersection number $I$ of an immersion in
terms of $t^{\mu\nu}$.
\beqa
\label{chi}
\chi&=&\frac{1}{4\pi}\int_\Si \ep^{ab}\p_a t^{\mu\nu}\p_b
t^{\mu\rho}t^{\nu\rho}\\
\label{inter}
I&=&-\frac{1}{16\pi}\int_\Si \sqrt{g}g^{ab}\p_a t^{\mu\nu}\p_b
{\tilde t}^{\mu\nu}
=\frac{1}{8\pi}\int_\Si \ep^{ab}\p_a t^{\mu\nu}\p_b
t^{\mu\rho}{\tilde t}^{\nu\rho}
\label{top}
\eeqa
where ${\tilde t}_{\mu\nu}=\half \ep^{\mu\nu\rho\si}t^{\rho\si}$.

The paper is organized as follows: in the first section we show that there is
an infinite energy barrier between instantons belonging to different topological
sectors of rigid string. This indicate the existence of instantons in each
topological sector of the model. In Sec.2 we derive basic equations,
while in the 
next section we discuss their solutions. Finally we comment on
other works devoted to the subject and state conclusions.

\section{Energy barrier between different instantons}

Before we go to the discussion of the instanton equations we shall show that 
any  action containing the rigidity has a minimum in each 
topological sector given by the Euler characteristic $\chi$ of $\Si$
and the self-intersection number $I$ of the immersion $X$. 
The considerations are valid for compact
surfaces only.

It is known that generic maps of a Riemann surface of genus $h$ to $R^4$ 
($X:\Si\to R^4$) are 
immersion. Immersions of given $\Si$ are classified, 
up to regular homotopies,
by the self-intersection number $I$ \cite{whitney1} (see also \cite{nfold} for a
brief review and some definitions). Hereafter we shall
identify both topological numbers $\chi,I$ with analytical 
expressions \refeq{chi} and 
\refeq{inter}, respectively.
If so the genus $h$ of $\Si$ is not really an invariant of continuous
deformations of $X$ but can acquire arbitrary values from metric singularities.
Similar behaviour characterizes $I$ what can be inferred from the similarity
of the 
expressions \refeq{chi} and \refeq{inter}. In the following we shall discuss
the latter case more thoroughly.  We shall construct a continuous family of maps 
$X_\a$ which will connect two
immersions with $I$ different by one. Thus the family will not be a regular
homotopy. $X_\a$ must go 
through a singularity i.e. a point where
the induced metric will vanish. We shall show that at this point the
rigidity is infinite. An action with rigidity  will separate different
topological sector of field configurations. Hence there must exist a minimum
of the rigidity for each $I$ and also for each $\chi$. 

Any map $X$ with 
given ($\chi,\,I$) can be locally deformed, by a homotopy which is not
regular, in such a way that  
$I$ will change by one. 
For a certain value of the deformation parameter, say $\a=0$,
the map $X_{\a=0}$ ceases to be
an immersion. The problem is to characterize singularities of 
$X_\a$ under such deformations. 
We shall parameterize
family $X_\a$ by $\a$ from neighborhood of zero $\a\in D^1$. 
Because deformations are local instead of 
considering the whole Riemann surface $\Si$ we take a 2d disc $D^2\subset\Si$. 
Thus the family  of discs is  a 
3d manifold $D^1\times D^2$. Maps $X_\a$ from  $D^2\subset\Si$ to
$R^4$ will be constructed as a composition of two maps: $X_\a=g\circ f_\a$,
 where $f_\a:D^2\to D^1\times D^2$ and $g:D^1\times D^2\to R^4$.
The first map $f$ must be non-singular i.e. it must be embedding 
because $X_\a$ must be immersion for all $\a\neq 0$.

In order to analyze singularities of $X_\a$ we must consider maps ${\tilde g}$
of $D^2$ 
together with parameter space $\a\in D^1$ into   
the 5-manifold $D^1\times R^4 $. The requirement is that the parameter space
is embedded into $D^1$ of $D^1\times R^4$. Hence $\p_\a {\tilde g}$ is never
zero.  
The generic singularities of such maps are well known
\cite{morin} to be cross-caps which in suitable coordinate system have the
form:
\beq
{\tilde g}:(t_1,t_2,x)\to (t_1,t_2,t_1 x,t_2 x,x^2).
\label{sing}
\eeq
The map has the line of self-intersections ${\tilde g}(0,0,x)=
{\tilde g}(0,0,-x)$ which terminates at the
singular point $x=0$. We must immerse  family ($f_\a$) of discs
$D^1\times D^2\to D^1\times R^4$ in such a way   
that it intersects (in 2 points) the line $t_1=t_2=0$ for
$\a<0$ and ceases to do it otherwise. Hence, for $a<0$,
$X_\a=g\circ f_\a$ is an immersion of  $D^2\subset\Si$ in $R^4$ with one
self-intersection point.  
$X_0=g\circ f_0$ ceases to be an immersion because it goes through singularity
point $(0,0,0)$ of ${\tilde g}$.

As $f_\a$ we consider a family of quadrics:
$f_\a(s,t)=\{s^2+t^2+\a,s,t\}$ in $D^1\times D^2$. 
It respects all requirement just 
imposed on the family of embedded surfaces. As $g$ we take the last four
components of the map (\ref{sing}) dropping the coordinate which corresponds to
an embedding of the deformation parameter ($\a\sim t_1$) in $D^1$. 
Thus $X_\a$ is:
\beq
\label{map}
X_\a(s,t)=g\circ f_\a(s,t)=\{s,(s^2+t^2+\a) t,s t,t^2\}
\eeq
As we expected, at $\a=0$ the image of $D^2$ under $X$ is singular i.e.
$\p X_0/\p t=0$ at $(s=t=0)$. For $\a\neq 0$ the map \refeq{map} is an immersion.
For $\a> 0$ it does not have self-intersection points ($I=0$).
For $\a< 0$ it has one self-intersection point: $X_\a(s=0,t=\sqrt{\a})=
X_\a(s=0,t=-\sqrt{\a})$ ($I=1$). The above arguments
show that (\ref{map}) is the generic form of maps with the desired properties.

Now we calculate the rigidity for such a family of maps.
The relevant formulae are \refeq{extc1}.
At $\a=0$ the density 
$\sqrt{g}g^{ab}g_{cd}K_a^{ib}K_b^{ia}$  diverges as: 
$4\pi/r^3 + O(1/r)$, where $r$ is the polar coordinate on $D^2\subset \Si$. 
Existence of the singularity means that the rigidity tends to 
infinity at $\a=0$ i.e. when $X$ ceases to be an immersion.
Thus {\it  
the rigidity  separates  configurations with
different self-intersection number by an infinite barrier}. 
Hence, we can expect a
minimum of an action with the rigidity for each topological sector of the
theory. 
This is the main conclusion of this part of the paper. Let us stress the local
aspect of the considerations, what implies its validity 
for an arbitrary target space-time.

In the rest of the
paper we shall be looking for these minima in terms of instantons. It will
appear that in some cases the minima do not exist if we bound considerations
to compact surfaces in $R^4$.

\section{Basic equations}

In this section we shall derive instanton equations. We recall that
the tensor $t^{\mu\nu}$
is the (Gauss) map $\Sigma\to G_{4,2}$, where $\Sigma$ is the Riemann
surface  and $G_{4,2}\equiv O(4)/(O(2)\times O(2))
=S^2\times S^2$ is the Grassman manifold of planes in 
$R^4$ ($\mu=0,1,2,3$). The product structure of  $G_{4,2}$ is related to the
fact that $t^{\mu\nu}$
splits into self-dual ($+$) and anti-self-dual ($-$) parts:
$t_\pm^{\mu\nu}\equiv t^{\mu\nu}\pm {\tilde t}^{\mu\nu}$. Both tensors assume
values in $S^2$ due to $t_\pm^{\mu\nu}t_\pm^{\mu\nu}=4$. In order to simplify
notation we introduce two vectors: $n_\pm^i=t_\pm^{0i}$ ($i=1\dots
3$)  which parameterize all components of $t_\pm^{\mu\nu}$ and respect 
$\vnp^2=\vnm^2=1$. There are associated topological
invariants $I_\pm$ which classify homotopy classes of maps\footnote{In the
case under consideration, homotopy classes of maps $\Sigma\to 
G_{4,2}$  are classified by their degrees. This follows from
the Pontryagin-Thom construction \cite{bredon}.} $\Sigma\to
G_{4,2}$. These are the degrees (winding numbers) of maps
$t_\pm^{\mu\nu}:\Si\to S^2$. 
Both topological invariants \refeq{top} can be expressed in terms of 
$I_{\pm}$.
\beq
\chi=I_+-I_-\quad I=\half(I_++I_-)
\label{char}
\eeq
where $I_{\pm}=\frac{1}{8\pi}\int_\Si \ep^{ab}\p_a{\vec n}_\pm(\p_b{\vec n}_\pm
\times {\vec n}_\pm$). We also note another useful
identity: 
\beq
I=-\frac{1}{16\pi}\int_\Si \sqrt{g}g^{ab}(\p_a\vnp\p_b\vnp-\p_a\vnm\p_b\vnm)
\label{ichar}
\eeq
which stems from (\ref{top}).
Using \refeq{char} and \refeq{ichar} we get
\beqa
\int_\Si\sqrt{g}g^{ab}\p_a t^{\mu\nu}\p_b t^{\mu\nu}
&=&\int_\Si\sqrt{g}g^{ab}(\p_a\vnp\p_b\vnp+\p_a\vnm\p_b\vnm
)\label{extc2}\\
&=&2\int_\Si\sqrt{g}g^{ab}\p_a\vnp\p_b\vnp+16\pi I
\label{idp}\\
&=&2\int_\Si\sqrt{g}g^{ab}\p_a\vnm\p_b\vnm-16\pi I
\label{idm}
\eeqa
In order to derive instanton equations we follow the standard route.
Let us write the inequalities:
\beq
\int_\Si \sqrt{g}g^{ab}(\p_a \vnp\pm \frac{\ep_a^{\;c}}{\sqrt{g}}
\p_c \vnp\times \vnp)(\p_b \vnp\pm
\frac{\ep_b^{\;d}}{\sqrt{g}}\p_d \vnp\times \vnp)\geq 0 
\eeq
which imply $\int_\Si\sqrt{g}g^{ab}\p_a\vnp\p_b\vnp\geq 8\pi |I_+|$.
They are saturated by the following instanton equations for the
self-dual part of $t^{\mu\nu}$ i.e. for  $\vnp$:
\beq
({+,\pm})\equiv\p_a \vnp\pm \frac{\ep_a^{\;c}}{\sqrt{g}}\p_c \vnp\times \vnp
=0 
\label{instp}
\eeq
There is a twin set of instanton equations for $\vnm$ i.e. for the
anti-self-dual 
part of $t^{\mu\nu}$. 
\beq
({-,\pm})\equiv\p_a \vnm\pm \frac{\ep_a^{\;c}}{\sqrt{g}}\p_c \vnm\times
\vnm =0 
\label{instm}
\eeq
As we shall see below, \refeq{instp} and \refeq{instm} are not
independent equations. This is obvious if one notices that $\vnp$ and $\vnm$ 
carry altogether the same degrees of freedom as $X$.
It follows that if one threats $\vnp$ and $\vnm$  as would be independent the
so-called 
integrability conditions appears. These will be discussed in the end of the
paper. Moreover one must realizes that  metric also depends only on the same
degrees of freedom. 

Hereafter we shall discuss relations between Eqs.(\ref{instp},\ref{instm}). 
If (\ref{instp}) holds then
$I_+\geq 0$ for
$({+,-})=0$ and $I_+\leq 0$ for $({+,+})=0$. 
On the other hand  if (\ref{instm}) holds then $I_-\geq 0$ for
$({-,-})=0$ and $I_-\leq 0$ for $({-,+})=0$.  
Let us check when two instanton equations can be respected
simultaneously. From (\ref{ichar}) we get 
$I=\half(-|I_+|+|I_-|)$. Confronting with
\refeq{char} we conclude that in this case $
|I_-|-I_-=I_+ +|I_+| $ must be respected.  All possible solutions to
this condition are listed below.  
\begin{enumerate}
\item $I_+>0$ implies $I_-< 0$. Instanton equations: $({-,-})=0,\;({+,+})=0$.
Below we shall show that, in fact  $({-,-})=0 \Leftrightarrow
({+,+})=0\Leftrightarrow \D X^\mu=0 $.
\item $I_+=0$ implies $I_-\geq 0$. Instanton equations are
$({+,-})=({+,+})=0$ and $(-,-)=0$. Due to the first point we get 
$\p_a t_+^{\mu\nu}=0$.
\item $I_-=0$ implies $I_+\leq 0$. Instanton equations $({-,+})=({-,-})=0$
and $(+,+)=0$. Due to the first point we get $\p_a t_-^{\mu\nu}=0$.
\item $I_+<0$ implies $I_-> 0$. Instanton equations:
$({+,-})=0,\;({-,+})=0$. Due to (\ref{char}) and $\chi\leq 2$, 
both equations can be
respected simultaneously only for  $I_+=1, I_-=-1$ (non self-intersecting 
sphere).
\end{enumerate}
We rewrite $({+,+})=0$ in terms of components of the stress tensor
$K^i_{ab}$. 
\beq
K^i_{ef}\le \d_{ij}\lb
-\d_a^e\frac{\ep^{\;f}_b}{\sqrt{g}}+\d_b^f\frac{\ep^{\;e}_a}{\sqrt{g}} \rb +
\ep_{ij}\lb \d_a^e\d_b^f+\frac{\ep^{\;e}_a\ep^{\;f}_b}{g}\rb\re=0
\label{mini}
\eeq
The l.h.s. of \refeq{mini} is equivalent to  $K^{i\;a}_a=0$ i.e. to $\D
X^\mu=0$.  Analogously one can show that $({-,-})=0 \Leftrightarrow \D
X^\mu=0$. Instantons respecting $\D X^\mu=0$ are called minimal.
Because the l.h.s. of  \refeq{extc1} is non-negative, minimal
instantons can not exist for Riemann surfaces of 
genus smaller than 2. There is one exception to this: torus with $I=0$ -
from  \refeq{extc1} we get $\p_a t^{\mu\nu}=0$ i.e. the ``torus'' is in fact 
degenerate to $R^2$.

We summarize this discussion noting that we obtained three families of instanton
equations: $({+,-})=0,\;({-,+})=0,\;({+,+})\equiv({-,-})=0$. 
Instantons considered in
\cite{polrig,wheater,rob} lies in the intersection of these families and
corresponds to the equations $\p_a t_\pm^{\mu\nu}=0$.  

It is useful to construct a map of all possible instantons on
the $(I,h)$ plane (here $h$ is the genus of the Riemann surface $\Si_h$). 
Minimal instantons respect  $I_+<0$ and  $I_-> 0$ thus leads to inequality
$|I|\leq h-1$; $({+,-})=0$  instantons respect $I_+\geq 0$ and from
\refeq{idp} we get 
$I+I_+=2I_++h-1\geq 0$; $({-,+})=0$ instantons respect $I_-\leq 0$ and  from
\refeq{idm}  
$I+I_-=2I_++1-h\leq $.  We also notice that instantons may 
exist for all possible $\chi$ and $I$.
Fig.1 summarizes the relation between different type of instanton equations.
\begin{figure}[t]
\label{pathnf}
\vspace{0.5cm}
\postscript{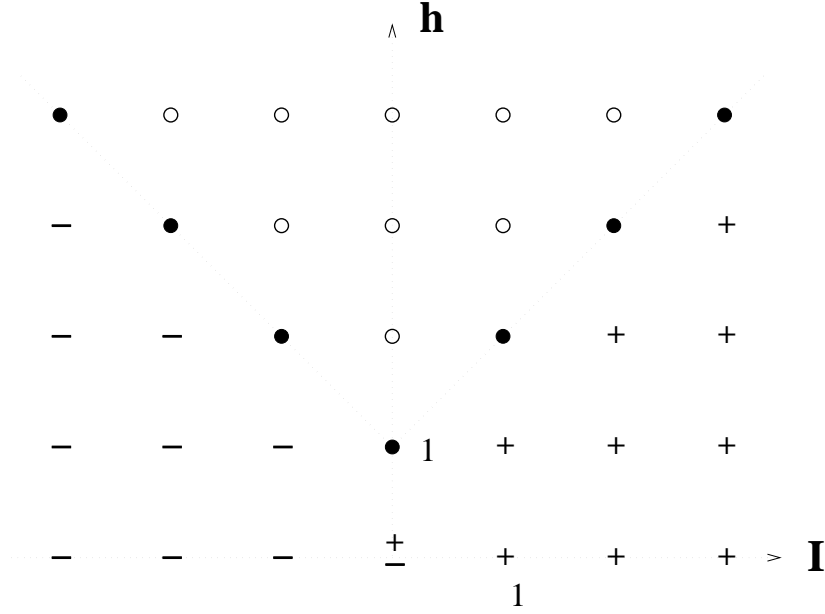}{0.67} 
\vspace{0.5cm}
\caption{ Rigid string instantons on the $(I,h)$ plane. Minimal instantons,
are denoted by empty circles, $({+,-})=0$ instantons by $+$'s, $({-,+})=0$
instantons by $-$'s, respectively.  
Instantons $\p_a \vn_\pm=0$ are denoted by full circles.
The solution found in this paper corresponds to the $\pm$ point.}
\end{figure}

\section{Solutions of instanton equations}
Apparently the problem of solving
Eqs.(\ref{instp},\ref{instm}) is very complicated. Despite this
some results are known.
On of the tools is 
the Gauss map of an immersion
\cite{osserman}. Let us recall some basic facts. The Gauss map of an immersion
$X:\Si\to R^4$ is defined to be the map $G:\Si\to G_{2,4}=S^2\times S^2$ i.e.
$G(z)$ gives tangent plane to the immersion at the point $X(z)$. 
For the conformal metric  $g_{ab}\propto \d_{ab}$ one can
identify  $G_{2,4}$ with a quadric in $CP^3$: $\sum_{\mu=1}^4 Z_\mu^2=0$, where
$Z_\mu$ are coordinates on $CP^3\subset C^4$. $Z_\mu$ is 
$\p X$ up to a $C$-number function $\Psi$ : $\p X^\mu=\Psi Z^\mu$.
Unfortunately not
every map $G:\Si\to G_{2,4}=S^2\times S^2$ can be a Gauss map of an immersion.
The so-called integrability conditions have to be respected \cite{osserman}. 
They originate from the fact that $\pb \p X^\mu$ must be orthogonal to 
$\p X^\mu$  
and real. Both conditions reads:
\beq
\pb \ln(\Psi)=-\frac{\pb Z^\mu {\bar Z}^\mu}{|Z|^2}, \quad
{\rm Im}\le \Psi\lb\pb Z^\nu ( \d^{\mu\nu}-
\frac{Z^\mu {\bar Z}^\nu}{|Z|^2})\rb\re=0
\label{second}
\eeq
There is a nice parameterization of $Z$
\beq
Z=\{ 1+f_+f_-,i(1-f_+f_-),f_+-f_-,-i(f_++f_-)\}, 
\label{param}
\eeq
where $f_i:\Si\to S^2$. 
From Eq.\refeq{second} one can derive the following 
integrability conditions \cite{osserman}:
\beqa
&&\frac{|\pb f_+|}{1+|f_+|^2}=\frac{|\pb f_-|}{1+|f_-|^2}
\label{con1},\\
&&{\rm Im}\le\pb\lb\frac{\p\pb f_+}{\pb f_+}-2\frac{\p f_+{\bar f}_+}{1+|f_+|^2}+
\frac{\p\pb f_-}{\pb f_-}-2\frac{\p f_-{\bar f}_-}{1+|f_-|^2}\rb\re=0
\label{con2}
\eeqa
Both conditions take relatively simple form when expressed in
terms of $({+,+}),({-,-})$: $|({+,+})|=|({-,-})|\; , \quad dA=0$
where $A=[\overline{(+,+)}\p ({+,+}) +\overline{(-,-)}\p
({-,-})]/|({+,+})|^2\;dz+c.c.$. We see that the first integrability condition
guarantee  the equality (\ref{idp})=(\ref{idm}). 
For minimal instantons the first condition is a
tautology while the second one looks singular.
There is a theorem \cite{osserman} which says that while the integrability
conditions  
(\ref{con1},\ref{con2}) are solved for appropriately regular maps we can
reconstruct 
the surface $X$ up to a 4d shift and a scale. 

It is worth to notice that the
integrability conditions posses symmetry 
groups. First of all it is the rotation group  $SO(4)\sim SO_+(3)\times
SO_-(3)$. 
\beq
f_\pm\to \frac{\a_\pm f_\pm+\b_\pm}{-{\bar \b}_\pm f_\pm +{\bar \a}_\pm}, 
\quad |\a_\pm|^2+|\b_\pm|^2=1,\quad
\a_\pm,\b_\pm\in C 
\label{so}
\eeq
Both conditions are also invariant under
(restricted) conformal transformations
performed on $f_+$ and $f_-$ simultaneously:
$f_\pm(z,\zb)\to f_\pm(g(z),\overline{g(z)})$.
This symmetry is the remnant of the reparameterization invariance of the
original theory. 

Below we shall shortly discuss solutions to the instantons equations and the 
above integrability
conditions. In the parameterization \refeq{param}:
\beqa
\vnp=(\frac{f_+{\bar f}_+-1}{1+|f_+|^2},\;i\frac{f_+-{\bar
f}_+}{1+|f_+|^2},\;\frac{f_++{\bar f}_+}{1+|f_+|^2})\non
\vnm=(\frac{f_-{\bar f}_--1}{1+|f_-|^2},\;i\frac{{\bar
f}_--f_-}{1+|f_-|^2},\;\frac{f_-+{\bar f}_-}{1+|f_-|^2})
\eeqa
so that we get 
\beq
({+,-})=0\Leftrightarrow\p f_+=0,\quad ({-,+})=0 \Leftrightarrow\p f_-=0
\label{holo}
\eeq
respectively. Hereafter we shall concentrate on the first of Eqs \refeq{holo}.
It can be easily solved: 
\beq
f_+=\eta_+\prod_{j=1}^{I_+}\frac{\zb-\ab_j}{\zb-\bb_j}.
\label{fplus}
\eeq
We solve the integrability conditions for the $I_+=1$ case.
Using conformal invariance we can
put $f_+=\zb$. Next we choose the ansatz for $f_-$:
${f}_-=\frac{a \zb+b}{c \zb+d},\quad ad-bc=1,\; a,b,c,d\in C$.
This is equivalent to an assumption that both Eqs.(\ref{holo}) are respected.
The second integrability condition holds identically.
The first
integrability condition gives  $d={\bar a},\; c=-{\bar b}$ thus setting
the solution on the $SO_-(3)$ manifold. Hence the whole moduli space of
solutions for $f_\pm$ consists of one point (up to irrelevant rotations of
space-time and
reparameterizations of the world-sheet). From (\ref{second}) 
we can determine
$\Psi$ : $\Psi=i\la/|Z|^2$, $\la\in R$.
Integrating $\p X=\Psi Z$ we get the immersion $X$
\beq
X-X_0=\frac{\la}{1+|z|^2} \{y,x,0,1\}
\label{sphere}
\eeq
The above is the sphere
$(X^0-X_0^0)^2+(X^1-X_0^1)^2+(X^3-X_0^3-\la/2)^2=\la^2/4$, $X^2-X_0^2=0$. 
The formula \refeq{sphere} gives 5-dimensional family of instantons.
In the forthcoming paper \cite{next} we show that this is really the most
general instanton 
family with $\chi=2,I=0$.   

Unfortunately for $I_+>1$ the situation is much more complicated. 
The simple ansatz
$f_-=\eta_-\prod_{j=1}^{m}\frac{\zb-\ab_j}{\zb-\bb_j}\prod_{k=1}^{m'}
\frac{z-\a_k'}{z-b_k'}$ appeared to be too restrictive and we were not able 
to find
any solutions to the integrability conditions. Definitely, different methods are
required \cite{next}.  

\section{Final comments}
Let us finally comment on other works concerning the rigid string instantons
and state conclusions.
 
Certain instanton equations for rigid  string were  proposed in
\cite{polrig} and farther elaborated in 
\cite{wheater,rob}\footnote{Surfaces constructed in \cite{rob} must be singular
i.e. they are not immersions. By a direct computations one can find that their
Euler  characteristic is (except one case) greater then 2.}. 
The considered equations were 
\beq
\p_a \vn_\pm=0
\label{polinst}
\eeq
One can  see  that they belong to the set  Eqs.(\ref{instp},\ref{instm})
restricted be the condition $I_\pm=0$. Eq. 
(\ref{char}) implies that for (\ref{polinst}) instantons $I=\pm\half\chi$
holds, so e.g. for the torus the 
equations can describe only the standard ($I=0$) immersion in $R^3$. 
Moreover (\ref{polinst}) implies also $\D X^\mu=0$. 
Hence no compact surface can be immersed in $R^4$ while
(\ref{polinst}) is respected. 

In the present paper we have shown that, contrary to 
\refeq{polinst}
general instanton equations (\ref{instp},\ref{instm}) can
have a representant for each value of $\chi$ and $I$. 
Not all of them can have compact
representant in non-compact  space-time $R^4$. Non-compactness of the
space-time makes some of the  
immersions to ``run away'' to infinity 
i.e. instantons become non-compact  and hard to control.
It is known that minimal
instantons $({+,+})=0$ 
can  not exist in $R^4$ \cite{osserman2,eells}. For $(+,-)$ and 
the twin $(-,+)$ family  
we have found
explicitly one compact instanton  
with topological numbers $\chi=2,I=0$. 
In the forthcoming paper we shall show that, in fact, all these instantons are
compact \cite{next}.

We want to stress
that despite this, the general arguments of 
Sec.1 shows that solutions to the instanton equations should 
exist for all possible topological sectors for compact space-times.
This subject goes beyond the scope of this
paper. 

We finish with few remarks concerning possible applications of
the rigid string instantons described in this paper. 
It is conceivable  that they may play 
prominent role in string description of gauge fields. For example, it is 
known that YM$_2$ in 1/N expansion is localized on surface-to-surface 
holomorphic and anti-holomorphic maps \cite{cmr} (see also \cite{horava}).
Four dimensional  
version of this construction was proposed in \cite{my,nfold}. 
Unfortunately, in this
case no definite set of maps was given. One may speculate that 
the rigid string instantons should be the appropriate maps. 
Work in this direction is in progress.

{\bf Acknowledgment}. I would like to thank  
A.Niemi for kind hospitality in Uppsala University where a part of this 
paper was prepared.

\end{document}